\documentclass[reprint,prm,superscriptaddress]{revtex4-1}
\usepackage{graphicx,color,charter}
\usepackage[version=4]{mhchem}

\begin{document}

\title{Computational screening of magnetocaloric alloys}

\author{Christina A. C. Garcia} 
\affiliation{Materials Research Laboratory, University of California, Santa Barbara, California 93106, USA}
\affiliation{John A. Paulson School of Engineering and Applied Sciences, Harvard University, Cambridge, 
Massachusetts 02138, USA}

\author{Joshua D. Bocarsly} 
\affiliation{Materials Research Laboratory, University of California, Santa Barbara, California 93106, USA}
\affiliation{Materials Department, University of California, Santa Barbara, California 93106, USA} 
\email{jdbocarsly@mrl.ucsb.edu}

\author{Ram Seshadri}
\affiliation{Materials Research Laboratory, University of California, Santa Barbara, California 93106, USA}
\affiliation{Materials Department, University of California, Santa Barbara, California 93106, USA} 
\affiliation{Department of Chemistry and Biochemistry, University of California, Santa Barbara, California 93106, USA}

\begin{abstract}
An exciting development over the past few decades has been the use of high-throughput computational screening 
as a means of identifying promising candidate materials for a variety of structural or functional properties. 
Experimentally, it is often found that the highest-performing materials contain substantial atomic site disorder. 
These are frequently overlooked in high-throughput computational searches however, due to difficulties in dealing 
with materials that do not possess simple, well-defined crystallographic unit cells. Here we demonstrate that 
the screening of magnetocaloric materials with the help of the density functional theory-based magnetic 
deformation proxy can be extended to systems with atomic site disorder. This is accomplished by thermodynamic 
averaging of the magnetic deformation for ordered supercells across a solid solution. We show that the highly 
nonmonotonic magnetocaloric properties of the disordered solid solutions Mn(Co$_{1-x}$Fe$_x$)Ge and 
(Mn$_{1-x}$Ni$_x$)CoGe are successfully captured using this method. 
\end{abstract}

\maketitle

\section{Introduction}

Recent advances in computing and automated materials science 
frameworks \cite{Curtarolo2012,Jain2013,Saal2013,Hautier2011,Hill2016,DePablo2019} have enabled 
high-throughput \emph{in silico} screening of crystalline solids aimed at identifying candidate materials 
for a variety of applications including structural materials \cite{Johannesson2002,Bligaard2003,Kirklin2016b}, 
battery electrodes \cite{Kirklin2013,Hautier2011}, thermoelectrics \cite{Yan2015}, 
photovoltaics \cite{Yu2012a,Fabini2019}, and magnetocalorics \cite{Bocarsly2017}, among many others. 
In these projects, automated density functional theory (DFT) calculations are performed on a large number of candidate
structures and compositions that have either been pulled from the literature or generated using a set of rules. 
Properties of interest are predicted from the results of these first principles calculations, often making use 
of a proxy: a simple quantifiable parameter that serves as an indicator of the more complex physical 
phenomenon \cite{Brgoch2013,Rondinelli2015,Bocarsly2017}. While this strategy has met with success and has expanded 
the breadth of materials systems under consideration for various applications, a major limitation is that these 
efforts have generally been limited to evaluating compounds with simple unit cells, and  without atomic site disorder 
(alloying). Consequently, alloyed and solid-solution materials are excluded from these searches, despite experiments
suggesting that the highest-performing materials for a variety of applications often come from these families.

The importance of screening compositionally disordered materials is especially apparent in the field of 
magnetocalorics, where many of the highest-performing materials rely on substantial unit cell disorder and 
nonstoichiometry for their remarkable properties, including 
(Mn,Fe)$_{2-\delta}$(P,Si) \cite{Tegus2002, Bruck2004, Thanh2008, Grebenkemper2018}, 
La(Fe,Si)$_{13}$H$_x$ \cite{Fujieda2002, Shen2009}, Gd$_5$(Si,Ge)$_4$  \cite{Pecharsky1997,Choe2000}, 
and a variety of substituted MnCoGe-based compounds \cite{Trung2010,Zhang2015a,Zhang2015b,Trung2010}.
In these materials, application of a magnetic field causes randomly oriented spins to align, reducing the 
entropy of the spin system. Alternating cycles of adiabatic and isothermal magnetization and demagnetization 
of a magnetocaloric can be used to drive a thermodynamic cycle and build an efficient magnetic heat 
pump \cite{Brown1976}. Such devices promise to provide an energy-efficient and environmentally-friendly alternative to 
conventional vapor-compression refrigeration and air conditioning \cite{Franco2012,Takeuchi2015}, which typically 
rely on hydrofluorochlorocarbons, which are now known to be associated with high global warming 
potential \cite{McLinden2017}. The primary metric used to quantify the performance of a magnetocaloric is the 
entropy change experienced by the material upon isothermal application of a magnetic $H$ field at a temperature 
$T$, $\Delta S_M(T,H)$. This parameter reaches its peak value near a magnetic transition temperature $T_c$, 
where the spins are most susceptible to an external field. An effective magnetocaloric should therefore show a large 
peak $|\Delta S_M(T,H)|$ at a useful temperature range.

For the high-performing magnetocaloric materials mentioned above, magnetic moments are strongly coupled to crystal 
structure, causing their magnetic transitions to couple to discontinuous changes in the crystal symmetry or 
lattice parameters. Such systems can show greatly enhanced (giant) magnetocaloric effects \cite{Pecharsky1997} 
around their first-order magnetostructural phase transitions. In fact, magnetostructural coupling can lead to 
an enhanced magnetocaloric effect even without this type of first-order transition 
present \cite{Bocarsly2019a, Dung2011, Franco2017, Davarpanah2019}. We previously introduced a simple 
DFT-based proxy for magnetostructural coupling known as the magnetic deformation \cite{Bocarsly2017} $\Sigma_M$
--- a stand-in for magnetostructural coupling ---  obtained through comparing the degree of lattice deformation 
between magnetic and nonmagnetic DFT structural optimizations. In systems where the inclusion of magnetism in 
the DFT calculation causes a large change in the optimized structure, we surmise that magnetostructural coupling 
must be strong. In a survey of reported magnetocalorics without substantial unit cell disorder, we found that 
$\Sigma_M$ correlates well with the experimental peak $\Delta S_M$ for transition metal-based compounds, 
both for materials with known first-order magnetostructural transitions and for those with no such 
transitions \cite{Bocarsly2017}. Consequently, $\Sigma_M$ can be used to computationally screen magnetic compounds
to identify promising magnetocalorics.

\begin{figure}
\centering
\includegraphics[width=.45\textwidth]{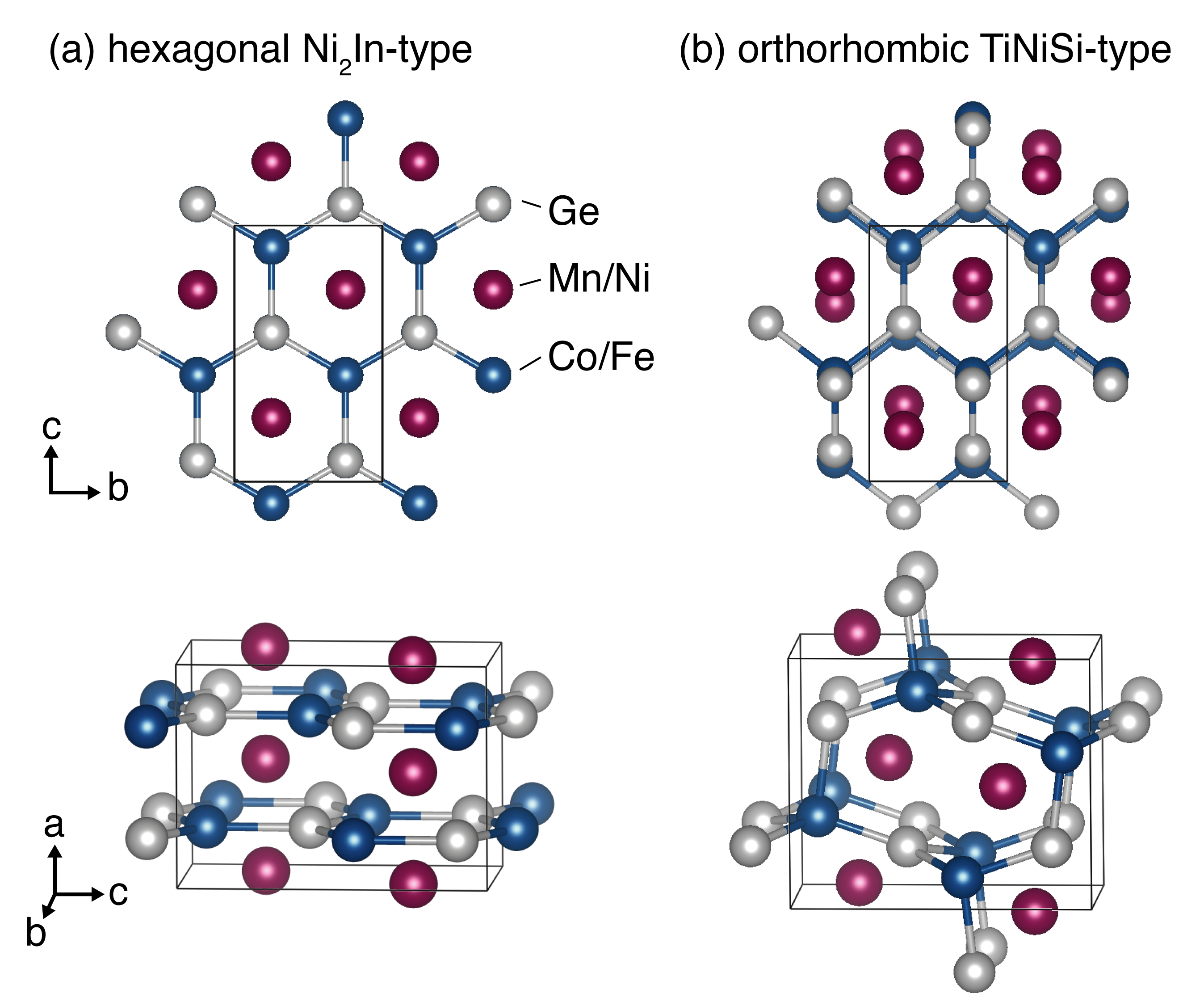}
\caption{(a) Hexagonal and (b) orthorhombic structure types of MnCoGe, depicted  in the orthorhombic 
setting.  The orthorhombic structure can be accessed from the hexagonal structure by a displacive phase transition 
involving corrugation of the honeycomb Co-Ge network.}  
\label{fig:structure}
\end{figure} 

While the application of the magnetic deformation proxy was previously limited to DFT-friendly compounds with 
no consideration of atomic site disorder, here we introduce a method to allow for the calculation of 
$\Sigma_M$ in disordered solid solutions. In order to accomplish this, we consider the test case of 
MnCoGe-based alloys. MnCoGe, an orthorhombic $Pnma$ compound with a TiNiSi-type structure, shows an intermediate 
peak $\Delta S_M$ of $-$6 J kg$^{-1}$ K$^{-1}$ for an applied field $H=5$~T \cite{Lin2006}. This effect is in 
agreement with the calculated value of $\Sigma_M$ = 1.93\% \cite{Bocarsly2017}. However, it was reported 
in 2010 that inclusion of just two or three percent boron (\emph{e.g.} MnCoGeB$_{0.02}$) in the material leads 
to a giant magnetocaloric effect with peak $\Delta S_M$ of up to $-$47.3\,J\,kg$^{-1}$\,K$^{-1}$ \cite{Trung2010}. 
This doped MnCoGe shows a coupled first-order magnetostructural transition, with a higher-symmetry hexagonal 
\ce{Ni2In} paramagnetic phase [Fig.\,\ref{fig:structure}(a)] transforming to a magnetic phase with a mixture of 
TiNiSi [Fig.\,\ref{fig:structure}(b)] and \ce{Ni2In} structures. The TiNiSi structure is described by a 
subgroup ($Pnma$) of the spacegroup of the \ce{Ni2In} structure ($P6_3/mmc$), and is formed by a displacive phase 
transition involving corrugation of the honeycomb Co-Ge lattice \cite{Landrum1998}, as illustrated in 
Fig.\,\ref{fig:structure}. Similar effects to those of boron-doping can also be realized with a number of other 
atomic substitutions, with giant magnetocaloric effects seen at disordered compositions including 
MnCoGeC$_{0.03}$ \cite{Trung2010}, Mn$_{0.9}$Ni$_{0.1}$CoGe \cite{Zhang2015a}, 
MnCoGe$_{0.95}$Ga$_{0.05}$ \cite{Zhang2015b}, and Mn$_{0.98}$CoGe \cite{Trung2010}. 

Here, we propose a method by which the magnetic deformation proxy $\Sigma_M$ can be used to screen compositionally 
disordered magnetic materials to identify promising magnetocaloric compositions. $\Sigma_M$ for a compositionally 
disordered material is calculated by taking a Boltzmann-weighted average of the individual $\Sigma_M$ values for 
enumerated ordered supercells of the disordered material. Using this technique, the qualitative magnetocaloric 
behavior of two solid solutions of MnCoGe are successfully reproduced: Mn(Co$_{1-x}$Fe$_x$)Ge \cite{Lin2006} and 
(Mn$_{1-x}$Ni$_x$)CoGe \cite{Zhang2015a}. In the first system, substitution of Fe for Co has has been shown to 
cause a modest increase in peak $-\Delta S_M$ at the intermediate composition $x$\,=\,0.2 \cite{Lin2006}. In the 
second, substitution of Ni for Mn has been shown to lead to a much larger increase in $-\Delta S_M$, with a giant 
magnetocaloric effect observed at $x$\,=\,0.1 \cite{Zhang2015a}. In both cases, we show that the highly nonmonotonic 
behavior of the solid solution is remarkably captured by the ensembled magnetic deformation calculations, with only 
minor deviations. We investigate the potential energy surfaces relevant to the DFT structural optimizations of 
individual supercell calculations for Mn(Co$_{1-x}$Fe$_x$)Ge and find that key cells experience double-well potentials 
with local minima at the hexagonal and orthorhombic structures of MnCoGe. This indicates that the ability 
of the structural optimization algorithm to traverse from one local minima to the other is an important 
consideration with regard to the results obtained from the magnetic deformation proxy calculations.

\section{Methods}

\subsection{Supercell enumeration}

For the solid solution systems studied, all possible orderings of the supercells up to a specified multiple of 
the volume of the 12 atom MnCoGe primitive cell ($Pnma$, TiNiSi structure) were enumerated. For the 
Mn(Co$_{1-x}$Fe$_x$)Ge system, we chose two times the primitive cell volume, allowing for $x$ increments of 1/8 
across the full composition range $x$\,=\,0 to $x$\,=\,1. For (Mn$_{1-x}$Ni$_x$)CoGe, supercells up to 3 times 
the primitive cell volume were considered from $x$\,=\,0 to 0.25, allowing for $x$\,=\,1/12 and $x$\,=\,1/6 
compositions to be probed in addition to $x$\,=\,0,~1/8~and~1/4. The Clusters Approach to Statistical Mechanics 
(CASM) code \cite{VanderVen2010,Thomas2013,Puchala2013} was used to enumerate these symmetrically distinct 
configurations and determine the multiplicity of each configuration. These ordered supercells may vary in cell 
shape and are not, in general, simple 2$\times$1$\times$1 or 3$\times$1$\times$1 stackings of the primitive cell. 
However, all cells do start with unit cell parameters and atomic positions consistent with the MnCoGe $Pnma$ symmetry, 
if the atom identity on the mixed site is ignored. Table~\ref{tbl:counts} lists the compositions for which we apply 
this method along with the number of supercell configurations generated and the maximum supercell size for each 
composition.

For the Mn(Co$_{1-x}$Fe$_x$)Ge system, a parallel set of supercells was also enumerated with the same unit cell 
orderings but with the atom positions and unit cell parameters adjusted to correspond to the symmetry of the 
Ni$_2$In-type hexagonal ($P6_3/mmc$) structure. The necessary transformation is possible for every 
supercell of the TiNiSi structure because $Pnma$ is a subgroup of $P6_3/mmc$ and therefore the \ce{Ni2In} structure 
type can always be expressed within a $Pnma$-compatible unit cell. 

\begin{center}
\begin{table}[hbt]
\begin{tabular}{|lrr||lrr|}
\hline
\hline
\multicolumn{3}{|c||}{MnCo$_{1-x}$Fe$_x$Ge} & \multicolumn{3}{c|}{Mn$_{1-x}$Ni$_x$CoGe} \\
\hline
$x$        & count & $V_{\rm{max}}$    & $x$   & count & $V_{\rm{max}}$ \\
\hline
0          & 1     & 1    & 0          & 1     & 1   \\
0.125      & 7     & 2    & 0.0833     & 9     & 3   \\
0.25       & 29    & 2    & 0.125      & 7     & 2   \\
0.375      & 41    & 2    & 0.1667     & 71    & 3   \\
0.5        & 58    & 2    & 0.25       & 184   & 3   \\ 
0.625      & 41    & 2    & \          & \     & \   \\
0.75       & 29    & 2    & \          & \     & \   \\
0.875      & 7     & 2    & \          & \     & \   \\
1          & 1     & 1    & \          & \     & \   \\
\hline
\hline
\end{tabular}

\caption{Compositions considered (labeled by $x$) and the number (count) of symmetrically distinct, ordered 
supercells with composition $x$ for the Mn(Co$_{1-x}$Fe$_x$)Ge and (Mn$_{1-x}$Ni$_x$)CoGe systems. For each 
composition, $V_{\rm{max}}$ is the volume of the largest supercells enumerated, in multiples of the primitive 
cell volume.}
\label{tbl:counts}
\end{table}
\end{center}

\subsection{Magnetic deformation}

For each enumerated cell, the magnetic deformation $\Sigma_M$ was calculated following the procedure given in 
Ref.~\citenum{Bocarsly2017}. The optimized structure for each configuration was acquired using density functional 
theory (DFT) with and without spin polarization. Calculations were performed using the Vienna \emph{ab initio} 
simulation package (VASP) \cite{Kresse1996} using the generalized gradient approximation (GGA) exchange-correlation 
functional as parameterized by Perdew, Burke and Ernzerhof \cite{Perdew1996, Blochl1994}. Spin-orbit coupling was 
not included. For each configuration, the spin-polarized relaxations were initialized with magnetic moments of 
3.0\,$\mu_B$ on each transition metal ion. 

Meshes for DFT calculations were automatically generated with the number of $k$-points set to 2500 divided by the 
number of atoms in the cell. Structural optimizations were performed using the conjugate gradient algorithm with an 
energy convergence criterion of 10$^{-3}$\,eV. The structural relaxations were run iteratively until the volume 
change between subsequent relaxations was less than 2\%. Once this convergence parameter was met, a final electronic 
optimization was performed for each enumeration while keeping the structure fixed. The Python packages 
\texttt{pymatgen} and \texttt{custodian} \cite{Ong2013} were used to automate, monitor, and analyze the VASP 
calculations.

Based on the obtained optimized structures, the magnetic deformation $\Sigma_M$ is calculated as the degree of 
lattice deformation (\%) \cite{Catti1985,delaFlor2016} between the DFT optimized nonmagnetic and magnetic structures. This is obtained by 
finding the transformation matrix between the two relaxed structures: 
$\textbf{P}=\textbf{A}^{-1}_{\text{NM}}\textbf{A}_{\text{M}}$, where \textbf{A}$_{\text{NM}}$ and 
\textbf{A}$_{\text{M}}$ are the lattice vectors of the nonmagnetic and magnetic relaxed unit cell, respectively. 
The Lagrangian finite strain tensor (which removes any rotational component of $\textbf{P}$) is then calculated 
as $\boldsymbol{\eta}=\frac{1}{2}(\textbf{P}^{\text{T}}\textbf{P}-\textbf{I})$, and the magnetic deformation is 
obtained as the root-mean-squared eigenvalues of $\boldsymbol{\eta}$:

\begin{equation}
\Sigma_M=\frac{1}{3}(\eta_1^2+\eta_2^2+\eta_3^2)^{1/2}\times 100\,\%.
\end{equation}

\noindent For the Mn(Co$_{1-x}$Fe$_x$)Ge system, in addition to the magnetic deformation calculated using only 
orthorhombic starting cells, a global $\Sigma_M$ was calculated for each cell based on the lowest energy nonmagnetic 
and the lowest energy magnetic structure obtained in either the run that started with the hexagonal structure or 
the run that started with the orthorhombic structure. 

Although it is well-established that DFT often fails to localize 3$d$ transition metal electrons enough to accurately model
the moments in magnetic intermetallics, we chose not to include any Hubbard $U$ correction terms in order to keep the 
calculations as simple (and generalizable) as possible, and to maintain compatibility with our previous work \cite{Bocarsly2017}
where it was found that $\Sigma_M$ performs well as a proxy for magnetocaloric effect across a diverse range of compounds
without the use of $U$. While we believe a GGA+$U$ approach could allow for the more faithful reproduction of magnetic
and structural ground states observed in experiment, this method increases computational cost and requires careful selection
of $U$ terms for each individual transition metal element in the compound, making it difficult to generalize to a high-throughput search. 

\subsection{Modeling disorder}

We consider the aggregate $\Sigma_M$ for a given composition labeled by $x$ to be determined by an ensemble of 
the ordered supercells. The aggregation may be done by a weighted average of the calculated $\Sigma_M$ for each 
ordering $i$ using the multiplicity $\Omega_i$ as the weight:

\begin{equation}\label{eqn:naive_avg}
\Sigma_{M,\rm{avg.}} = \frac{\sum\limits_i {\Omega_i \Sigma_{M,i}}}{\sum\limits_i{\Omega_i}}\,.
\end{equation}

\noindent A more complete picture, however, considers the calculated energy of each enumeration, considering that 
low energy states are more likely to be present in a true sample of a disordered alloy. To approximate this, we 
define the Boltzmann weight of a configuration $i$ with composition $x$ as

\begin{equation}\label{eqn:boltz_weight}
w_i=\Omega_i\exp{\left(\frac{E_i-E_0}{k_BT}\right)}
\end{equation}

\noindent such that the Boltzmann-weighted average $\Sigma_M$ is

\begin{equation}\label{eqn:boltz_avg}
\Sigma_{M,\rm{Boltzmann}} = \frac{\sum\limits_i{w_i\Sigma_{M,i}}}
{\sum\limits_i{w_i}}\,.
\end{equation}

\noindent Here, $E_i$ is the spin-polarized energy of supercell $i$, expressed \emph{per} the maximum supercell 
size (\emph{i.e.} in units of eV \emph{per} 24 or 36 atoms). $E_0$ is the energy of the lowest-energy enumeration 
for the composition $x$, and $k_B$ is the Boltzmann constant. The temperature $T$ was set to 300\,K. In addition, 
we also tested setting the temperature to the preparation temperatures of the alloys (around 1000\,K), and this 
did not dramatically change the presented results.

In addition to enumerating small supercells, we also tried calculations of $\Sigma_M$ on special quasirandom structures, a different method commonly used for DFT modeling of alloys \cite{Zunger1990,vandeWalle2013}. In this method, an alloy composition is modeled by a single large supercell (here, 48 atoms) with occupation of the atomic sites chosen so as to match the near-neighbor correlations of the true infinite disordered compound as well as possible. Unfortunately, this method was not as successful as the supercell enumeration method for the tasks investigated presently. For a discussion of these calculations, the reader is directed to the Supplemental Material. 

\subsection{Transition paths}

In order to investigate the potential energy surfaces which control the DFT structural relaxations used to 
calculate $\Sigma_M$, we performed transition path calculations on a few selected atomic supercells of 
Mn(Co$_{0.75}$Fe$_{0.25}$)Ge between their hexagonal and orthorhombic structures. Lattice parameters and atom 
positions of structures along the path are interpolated between the end members, which are the relaxed hexagonal 
($d = 0$) and orthorhombic ($d = 1$) structures. The energies of structures along this path were calculated 
without structural relaxation. 

\section{Results and Discussions}

\begin{figure}
\centering
\includegraphics[width=.45\textwidth]{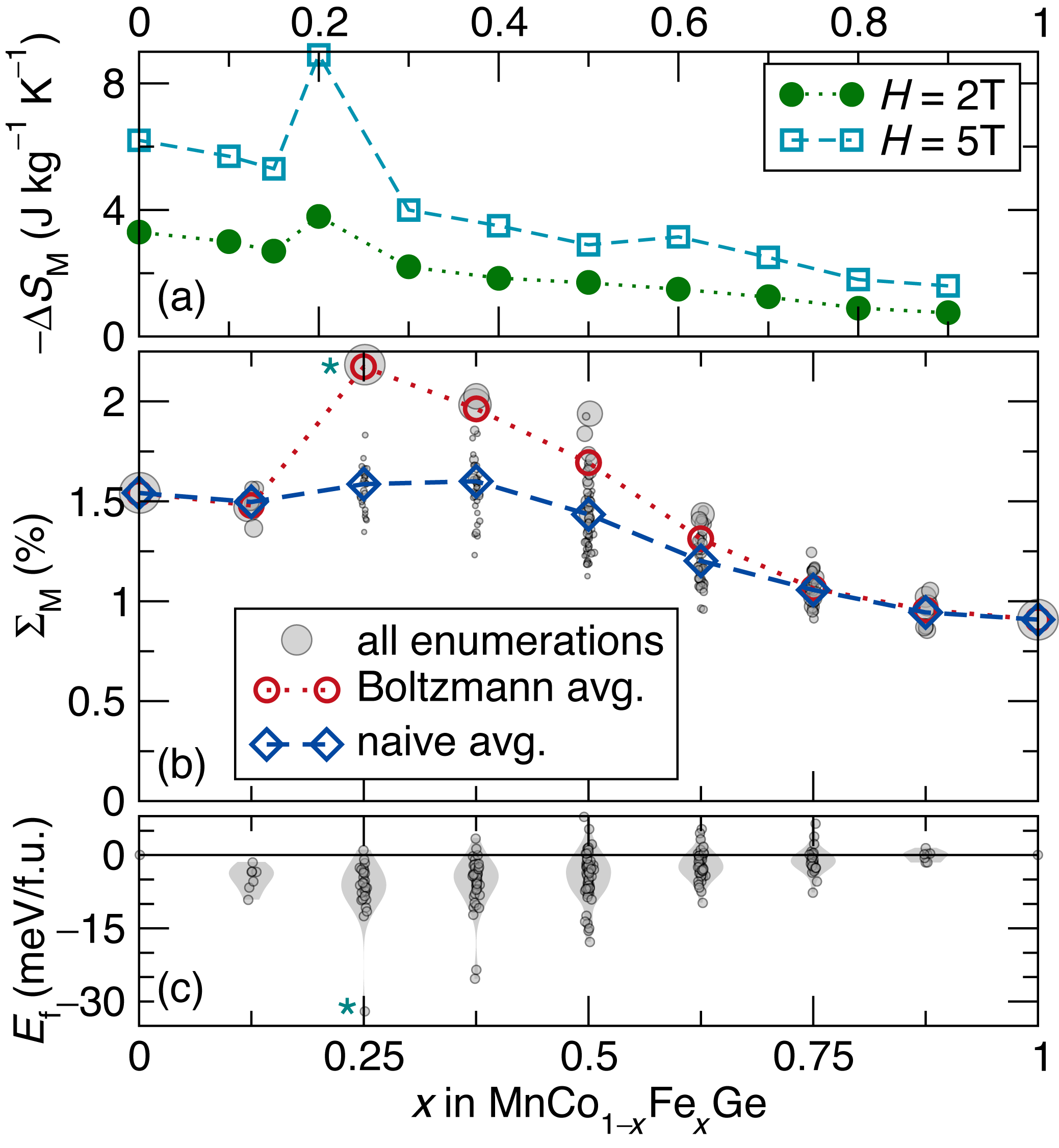}
\caption{Comparison of (a) peak $\Delta S_M$ values as measured by Lin \emph{et al.} for applied fields of 2\,T and 5\,T \cite{Lin2006} and (b) calculated 
$\Sigma_M$ \emph{vs.} $x$ for Mn(Co$_{1-x}$Fe$_x$)Ge. In (b), each gray circle represents a single enumerated 
cell, with the area of each circle proportional to its Boltzmann weight as calculated in 
equation~\ref{eqn:boltz_weight}. Both Boltzmann (equation~\ref{eqn:boltz_avg}) and naive 
(equation~\ref{eqn:naive_avg}) averages of $\Sigma_M$ for composition $x$ are overlaid. (c) Energy of 
formation \emph{vs.} $x$ for each cell. The asterisk indicates the cell indicated as cell $B$ in 
Fig.\,\ref{fig:landscapes}. The gray shaded areas (violin plot) visualize the distribution of the supercell energies. 
} 
\label{fig:MnCoFeGe_ortho}
\end{figure}

Experimental peak $-\Delta S_M$ values and computed $\Sigma_M$ data for the Mn(Co$_{1-x}$Fe$_x$)Ge system are 
shown in Fig.\,\ref{fig:MnCoFeGe_ortho}. MnCoGe and MnFeGe are both ferromagnets, and the full solid solution 
between them can be prepared experimentally \cite{Lin2006}. This solid solution features a transition 
from the orthorhombic $Pnma$ structure of MnCoGe at $x<$\,0.2 to the hexagonal $P6_3/mmc$ structure of MnFeGe 
at $x>$\,0.2 \cite{Lin2006}. Across this series, peak $-\Delta S_M$ decreases as $x$ increases, except for at 
the phase boundary ($x$\,=\,0.2), where a peak in $-\Delta S_M$ reaching 9\,J\,kg$^{-1}$K$^{-1}$ for an applied 
field of 5\,T is observed. Figure~\ref{fig:MnCoFeGe_ortho}(c) shows the energies of the individually enumerated 
supercells relative to the energies of the corresponding mixture of MnCoGe and MnFeGe. Many orderings across the 
full compositional range show negative formation energies, consistent with the experimental observation that the 
solid solution forms and does not phase segregate. As seen in Fig.\,\ref{fig:MnCoFeGe_ortho}(b), the calculated 
$\Sigma_M$ values for individual ordered cells span a range of values, from about 0.75\% to 2.25\%. The simple 
average of these $\Sigma_M$ values somewhat follows the experimental trend of a general decrease in $\Delta S_M$ 
with increasing $x$ interrupted by a peak near the middle of the compositional range. However, the position and 
magnitude of the peak in $\Sigma_M$ are far off from the experimental results, and therefore the correspondence 
between computation and experiment is poor. On the other hand, the Boltzmann-weighted average gives an excellent 
qualitative match, with a maximum $\Sigma_M$ at $x$\,=\,0.25, the closest computed composition to the peak in 
the experimental data ($x$\,=\,0.2). The peak in $\Sigma_M$ is broader than that seen in 
the $\Delta S_M$ data; however, the qualitative match is remarkable given the simplicity of the computational model 
and the many variables involved in the experimental preparation and measurement of a magnetocaloric material.

\begin{figure}
	\centering
	\includegraphics[width=.45\textwidth]{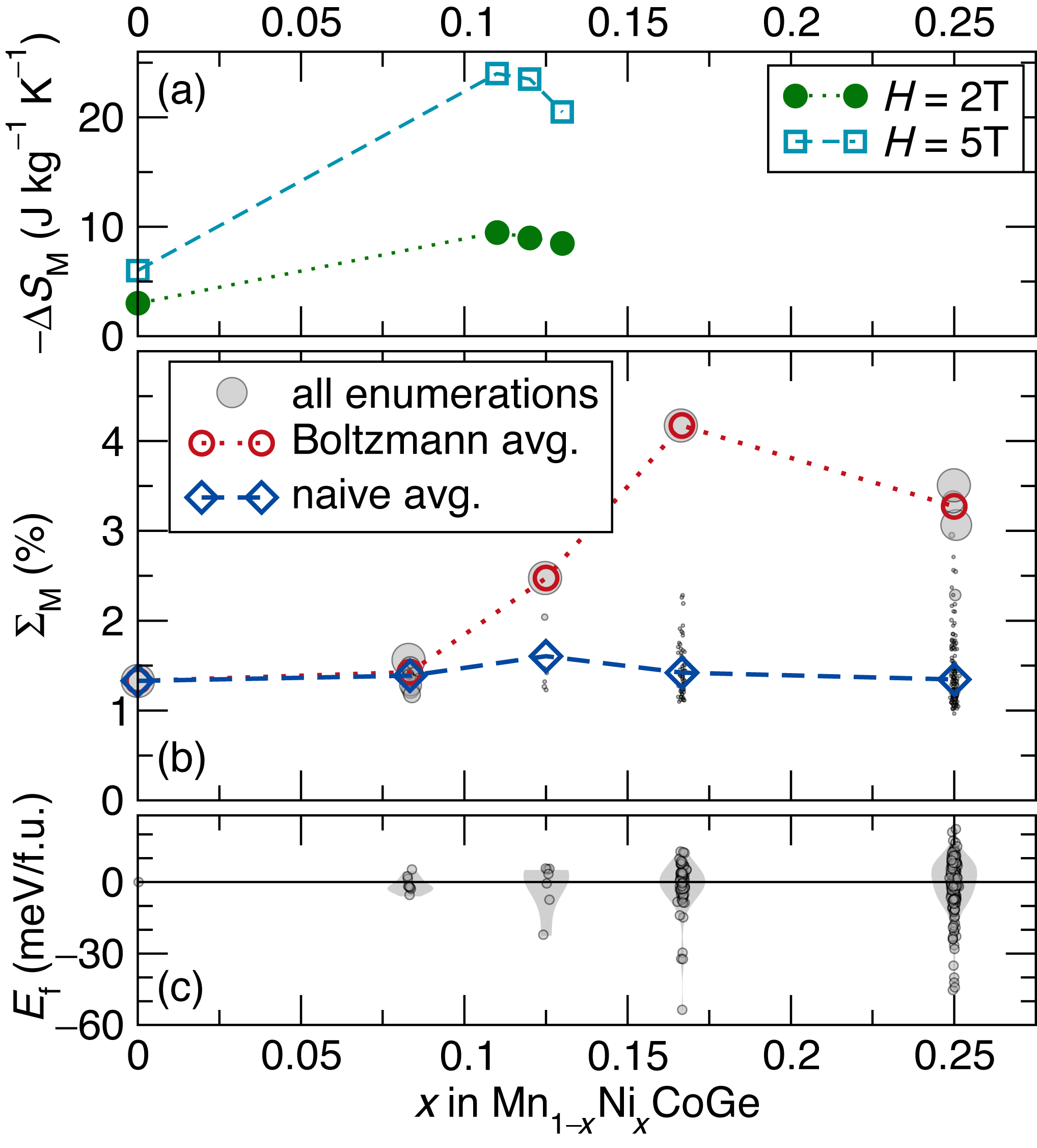}
	\caption{Comparison of (a) experimental peak $\Delta S_M$ values as measured by 
         Zhang \emph{et al.} \cite{Zhang2015a} and (b) calculated $\Sigma_M$ for (Mn$_{1-x}$Ni$_x$)CoGe ($x \leq 0.25$). 
         (c) Energies of formation \emph{vs.} $x$ for the enumerated cells. Refer to the 
          Fig.\,\ref{fig:MnCoFeGe_ortho} caption for additional definitions.}  
	\label{fig:MnNiCoGe}
\end{figure}

Figure~\ref{fig:MnNiCoGe} shows the same analysis for a different solid solution of MnCoGe, the (Mn$_{1-x}$Ni$_x$)CoGe 
system. In this case, introduction of a small amount of Ni ($\approx$ 11\%) has been found to result in a giant 
magnetocaloric effect with peak $-\Delta S_M$ reaching 24\,J\,kg$^{-1}$K$^{-1}$ for an applied field of 
5\,T \cite{Zhang2015a}. As in the Mn(Co$_{1-x}$Fe$_x$)Ge system, this is due to the coincidence of magnetic 
and structural transitions, \emph{i.e.} due to a first-order magnetostructural transition, observed for samples 
with $x$ between 0.08 and 0.12 (for $x$\,$<$\,0.08 and $x$ slightly greater than 0.12, the structural 
and magnetic transitions occur at different temperatures). As alloying across the whole composition 
space 0\,$\leq$\,$x$\,$\leq$\,1 has not yet been reported, in order to set a reasonable limit to the computational 
cost, $\Sigma_M$ was calculated only for $x \leq$ 0.25 for this system (Fig.\,\ref{fig:MnNiCoGe}). For this 
compositional range, the calculations presented here reproduce the experimental $\Delta S_M$ reports with a similar 
level of success as the study of the Mn(Co$_{1-x}$Fe$_x$)Ge system discussed above. While the maximum $\Sigma_M$ 
is slightly misaligned from the experimental largest $-\Delta S_M$ ($x = 0.167$ \emph{vs.} $x = 0.11$, respectively), 
the qualitative shape and the magnitude of the $\Sigma_M$ curve matches nicely to the experiment.

A direct comparison of the two systems under study reveals that the maximum $\Sigma_M$ is 2.1 times larger in the 
(Mn$_{1-x}$Ni$_x$)CoGe system than in the Mn(Co$_{1-x}$Fe$_x$)Ge system. Similarly, the ratio of the maximum 
Boltzmann averaged $\Sigma_M$ values is 2.7. Even without experimental references, a computational screen comparing 
these two systems would correctly conclude that (Mn$_{1-x}$Ni$_x$)CoGe is a more promising candidate system of 
experimental study. Furthermore, such a conclusion would be reached even if we had only considered supercells of 
up to size $V_{\rm{max}} = 2$ (24 atoms) for both systems. While the peak in $\Sigma_M$ at $x=0.167$ (1/6) in 
(Mn$_{1-x}$Ni$_x$)CoGe would not have been captured, the $\Sigma_M$ at $x=0.25$ is still large enough relative 
to any values in the Mn(Co$_{1-x}$Fe$_x$)Ge system to suggest that Ni is a more interesting dopant. A comparison of $\Sigma_M$ 
and $\Delta S_M$ values for the two systems on the same scale may be found in Supplemental Material Fig.~S3.

\begin{figure}
	\centering
	\includegraphics[width=.45\textwidth]{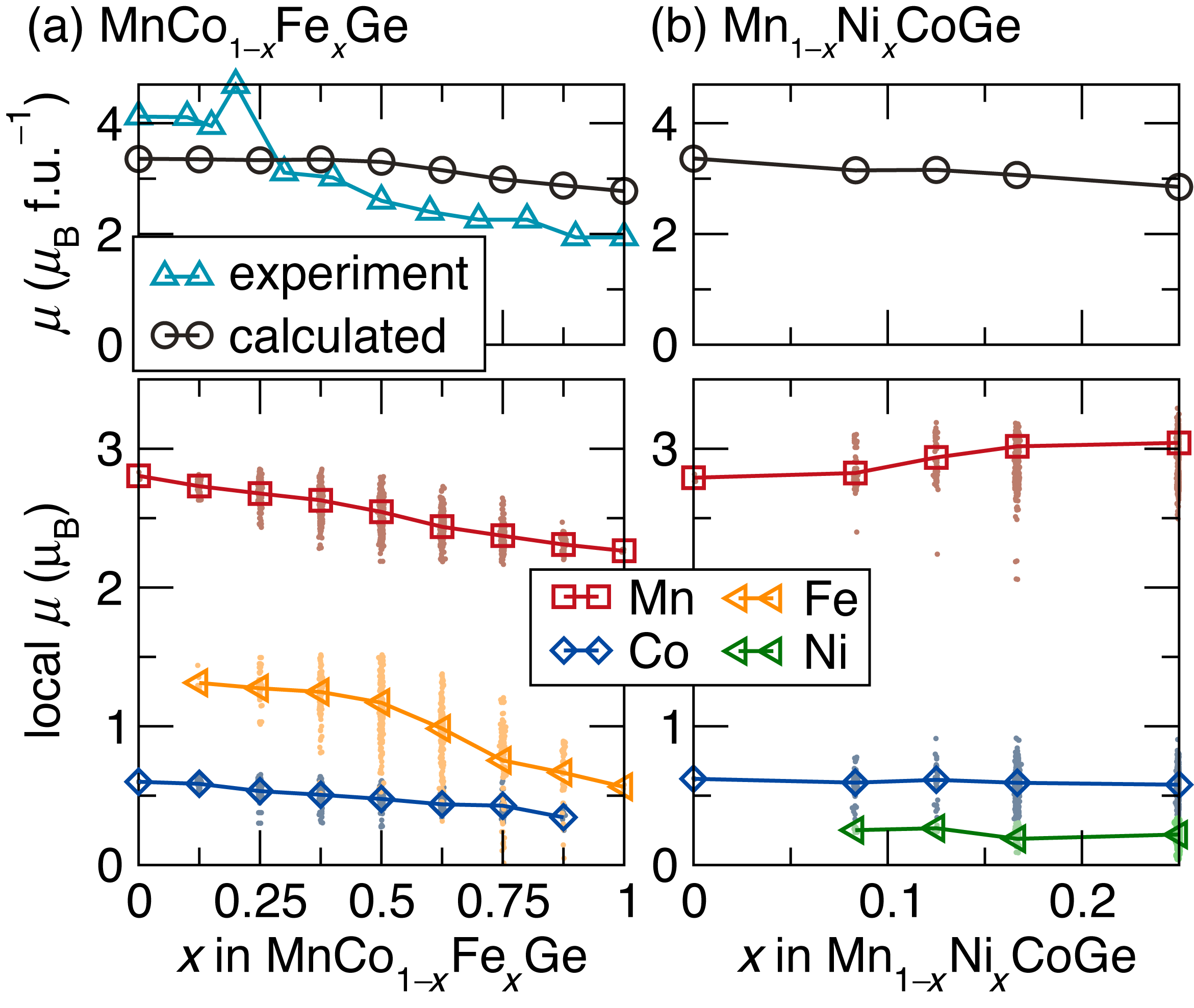}
	\caption{Total (top) and local (bottom) evolution of the DFT magnetic moments for (a) Mn(Co$_{1-x}$Fe$_x$)Ge 
	and (b) (Mn$_{1-x}$Ni$_x$)CoGe. All of the individual transition metal local moments from individual enumerated
	cells are shown as small dots, while the Boltzmann-averaged local and total moments are shown as larger symbols 
	connected by lines. For Mn(Co$_{1-x}$Fe$_x$)Ge, experimentally measured 5\,K saturated total magnetic moments
	from Lin \emph{et al.} \cite{Lin2006} are plotted for comparison.}  
	\label{fig:moments}
\end{figure}

In addition to $\Sigma_M$ and cell energy, we obtain information from our calculations about the evolution of magnetic moments
in these solid solutions. Figure~\ref{fig:moments} shows the DFT total moment (top) and projected local moments
(bottom) for Mn(Co$_{1-x}$Fe$_x$)Ge and (Mn$_{1-x}$Ni$_x$)CoGe. As with the ensembled $\Sigma_M$ calculations, the 
total and local moments shown are a Boltzmann-weighted average of all of the moments in all of the enumerated supercell 
calculations. However, in this case, there are not substantial differences between Boltzmann-weighted and simple averages. In the case of
 Mn(Co$_{1-x}$Fe$_x$)Ge, a comparison to the reported experimental saturated moments at 5\,K is included \cite{Lin2006}, while
 we were unable to find such data for (Mn$_{1-x}$Ni$_x$)CoGe. From this comparison, we see that the calculated moment
 is underestimated for MnCoGe and overestimated for MnFeGe. As discussed in the methods section, it is likely that a more 
 faithful reproduction of the experimental moments would require a GGA+$U$ approach. However, the general trend of 
 decreasing moment with increasing Fe content is captured by our calculations, and we can therefore use these results to draw insight into
 the local magnetic moment evolution. As Fe atoms are substituted in, they are found to hold a larger local moment than 
the Co atoms they replace (about 1\,$\mu_B$ \emph{vs.} 0.5\,$\mu_B$). However, at the same time, the large Mn moments decrease in 
magnitude with increasing $x$. The net effect is a decrease in total moment with $x$. In (Mn$_{1-x}$Ni$_x$)CoGe, a decrease
in total moment with $x$ is also predicted; however, in this case, the decrease is driven by the replacement of high-moment Mn atoms
(3\,$\mu_B$) with Ni atoms which have very small moments (about 0.25\,$\mu_B$).

Returning to the $\Sigma_M$ results, we noticed that for the compositions showing maximal Boltzmann-averaged 
$\Sigma_M$ ($x = 0.25$ for Mn(Co$_{1-x}$Fe$_x$)Ge and $x = 0.175$ for (Mn$_{1-x}$Ni$_x$)CoGe), the Boltzmann averages 
are dominated by a single enumerated cell which has significantly lower energy than the rest of the cells. For 
example, in Fig.\,\ref{fig:MnCoFeGe_ortho}(c), the energy of formation for all of the enumerated cells is plotted 
\emph{versus} $x$. At $x$\,=\,0.25, the cell marked with an asterisk is about 20\,meV\,f.u.$^{-1}$ lower in energy 
than all the other cells, and therefore contributes dominantly to the Boltzmann-averaged $\Sigma_M$.  This special 
unit cell also exhibits a larger $\Sigma_M$ than any of the other enumerations, and as a result this single cell 
is entirely responsible for the peak at $x$\,=\,0.25 in the Boltzmann-averaged $\Sigma_M$. Inspection of the 
calculations for this special cell revealed that the magnetic structural optimization resulted in a cell with 
atom positions consistent with the hexagonal structure [Fig.\,\ref{fig:structure}(a)], while the nonmagnetic structural 
optimization stayed in the orthorhombic structure [Fig.\,\ref{fig:structure}(b)] with which the calculation was 
initialized. For the other enumerations at $x$\,=\,0.25, both the magnetic and nonmagnetic unit cells remained in 
the orthorhombic structure.

\begin{figure}
	\centering
	\includegraphics[width=.45\textwidth]{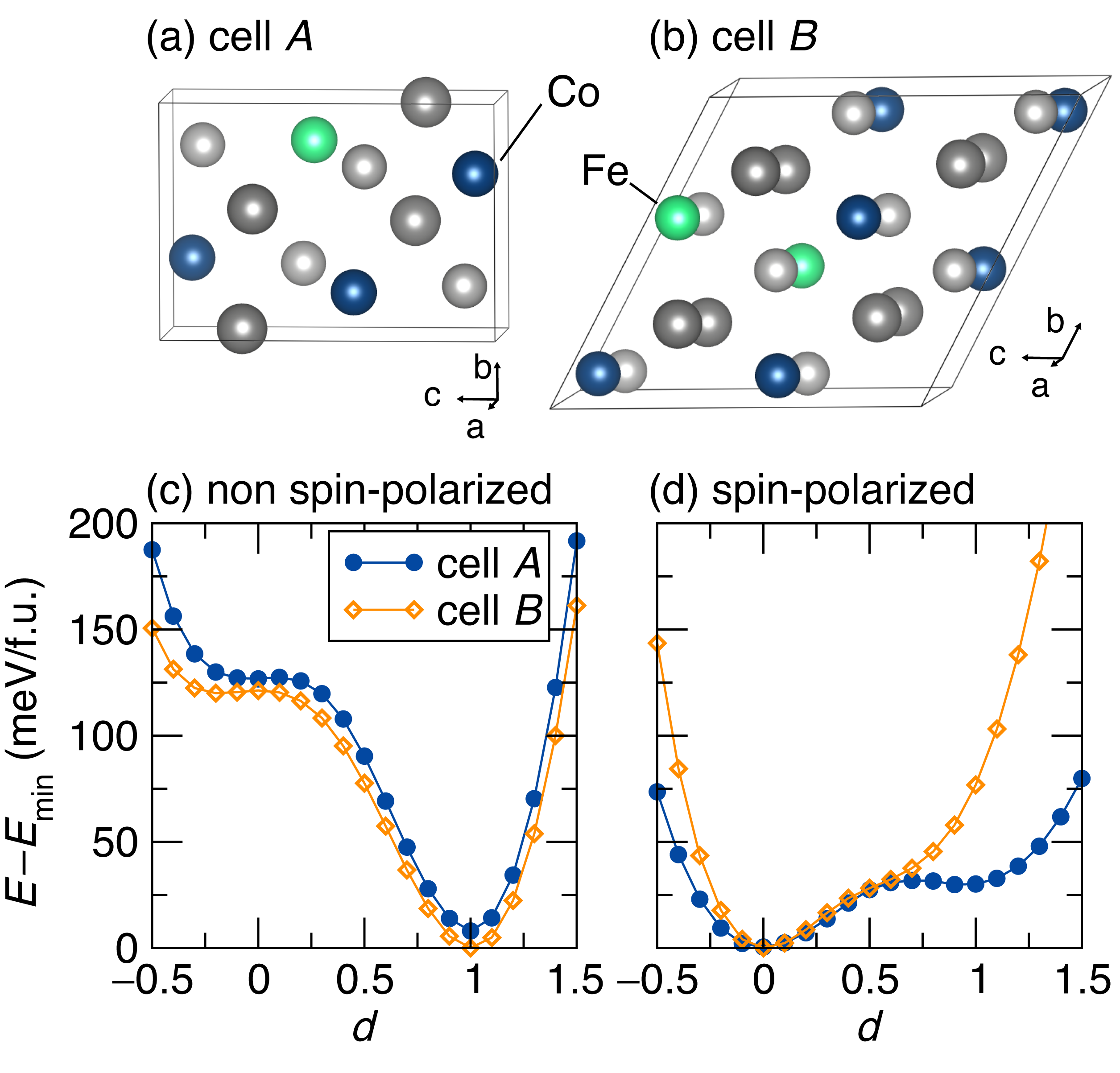}
	\caption{Transition path energies between the hexagonal (MnFeGe-like, $d$\,=\,0) and orthorhombic 
(MnCoGe-like, $d$\,=\,1) structures of two different enumerated configurations of MnCo$_{0.75}$Fe$_{0.25}$Ge, 
(c) without and (d) with spin polarization. (a) Cell \emph{A} shows a transition path landscape which is 
characteristic of that experienced by the majority of the enumerated cells, which remain orthorhombic after both 
nonmagnetic and magnetic structural relaxations. (b) Cell \emph{B} is the special cell marked by an asterisk 
in Fig.\,\ref{fig:MnCoFeGe_ortho}, which transformed to the hexagonal structure during the spin-polarized relaxation.}
	\label{fig:landscapes}
\end{figure}

To understand this, we turned to calculations of the transition path energies between the hexagonal and orthorhombic 
structures of two representative enumerated cells with $x$\,=\,0.25: cell $A$ is a cell that stayed in the 
orthorhombic structure for both magnetic and nonmagnetic optimizations, and cell $B$ is the special cell that changed 
structures during the magnetic optimization. For each cell, Fig.\,\ref{fig:landscapes} shows the energies of the 
transition paths with and without spin-polarization as functions of the fractional hexagonal distortion $d$, with 
$d = 0$ corresponding to the hexagonal structure, and $d = 1$ to the orthorhombic structure. Interestingly, for both 
cells, the nonmagnetic calculation shows a global minimum at the orthorhombic structure while the magnetic calculation 
shows a global minimum at the hexagonal structure. As the enhanced magnetocaloric effect around $x$\,=\,0.2 
in Mn(Co$_{1-x}$Fe$_x$)Ge is believed to arise from coupling of the magnetic transition to a structural transition, 
it is interesting to note that the inclusion of magnetism in the DFT calculation changes the predicted structural 
ground state. However, it is also important to note that the nonmagnetic DFT calculation should not be considered a 
realistic model for the true high-temperature paramagnetic state. 

The transition path energies without spin polarization look qualitatively similar for cell $A$ and cell $B$, 
with a very shallow local minimum at the hexagonal structure and a global minimum at the orthorhombic structure. 
In contrast, with spin polarization, greater differences between the two cells are evident. Cell $A$ exhibits 
a double well potential with a barrier between the wells, while cell $B$ has no barrier to relaxation into the 
global minimum hexagonal structure. As the optimizations used to calculate $\Sigma_M$ were initialized with an 
orthorhombic starting configuration, cell $A$ relaxed into the orthorhombic local minimum, while cell $B$ was 
able to relax into the global minimum structure. As a result of this feature of its potential energy surface, 
the DFT calculations on cell $B$ result in a lower energy and larger magnetic deformation than all other cells 
enumerated at this composition. We can therefore conclude that the effectiveness of the magnetic deformation proxy 
in identifying the extremal magnetocaloric composition in this system is driven by the ability to conveniently 
identify a potential energy surface with competing structural ground states whose energies are coupled to the 
system magnetism, and which has low barriers to relaxation from one state to the other. These features are consistent 
with the thermodynamic conditions necessary for a first-order magnetostructural transition leading to an enhanced 
magnetocaloric effect.

\begin{figure}
	\centering
	\includegraphics[width=0.45\textwidth]{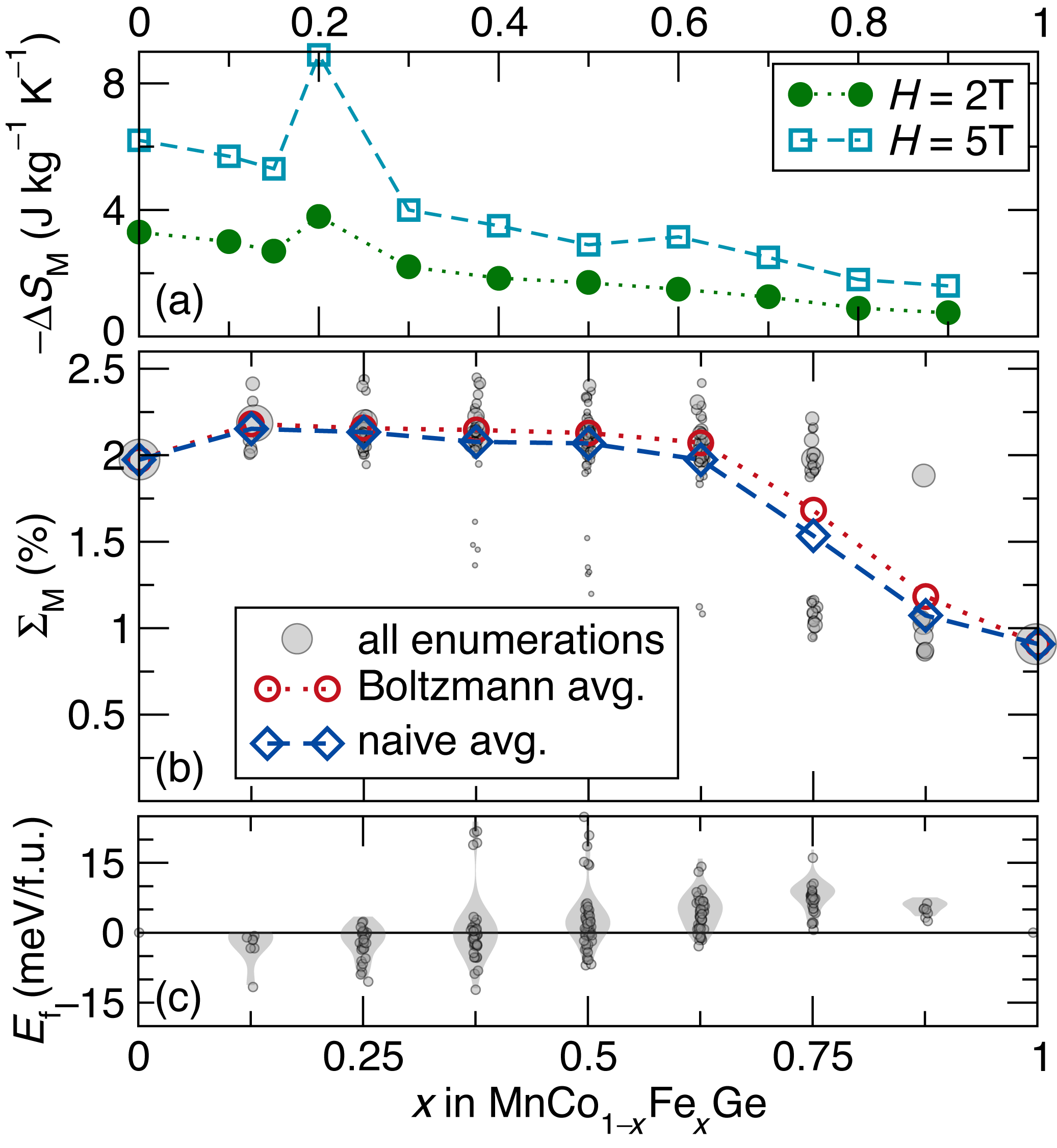}
	\caption{Comparison of peak $\Delta S_M$ values as measured by Lin \emph{et al.} \cite{Lin2006} (a) and 
calculated $\Sigma_M$ \emph{vs.} $x$ (b) for Mn(Co$_{1-x}$Fe$_x$)Ge using global $\Sigma_M$ values computed 
from calculations starting from both the orthorhombic and hexagonal structures, as discussed in the text. 
Please refer to the Fig.\,\ref{fig:MnCoFeGe_ortho} caption for definitions. }  
	\label{fig:MnCoFeGe_all}
\end{figure}

Based on this analysis, we proceed to consider what role the incomplete structural relaxations in cells like cell 
$A$ played in the evaluation of the overall $\Sigma_M$. To address this, a parallel set of DFT calculations was 
run with the enumerated supercells initialized in the hexagonal structure, instead of the orthorhombic structure. 
The nonmagnetic and magnetic structures used in calculating $\Sigma_M$ were then each taken from the calculation
that reached a lower energy state. The results are shown in Fig.\,\ref{fig:MnCoFeGe_all}, which can be 
compared to Fig.\,\ref{fig:MnCoFeGe_ortho} in which only the orthorhombic initialization was considered. Providing 
an alternate path to relaxation for each cell increases the likelihood that the global energetic minimum is reached 
for each of the nonmagnetic and magnetic optimizations. As a result, more cells change structure type between the 
nonmagnetic and magnetic unit cells and the $\Sigma_{M,i}$ values increase. Both the Boltzmann and simple 
weighted averages of $\Sigma_M$ are seen to increase for all $x$\,$<$\,0.8 in the Mn(Co$_{1-x}$Fe$_x$)Ge system 
such that the strong peak seen before at $x$\,=\,0.2 for the Boltzmann-averaged $\Sigma_M$ is smeared out and 
the qualitative match with the experimental data is weakened. Nevertheless, the composition and magnitude of the 
peak in $\Sigma_M$ remains very similar to the original calculations. Furthermore, this approach results in the 
Boltzmann and naive averages converging to nearly the same values for each composition. Therefore, if using this 
strategy to screen magnetocaloric systems, one does not necessarily need to exhaustively perform DFT calculations on 
all possible enumerations; rather, reasonably accurate results could be obtained by averaging together a small 
number of $\Sigma_{M,i}$ values (\emph{i.e}, 5 -- 10 cells) for each composition $x$. 

\section*{Summary and Conclusions}

In this work, we introduce a method for screening experimental magnetocaloric behavior in disordered compounds 
(alloys) which employs the magnetic deformation proxy $\Sigma_M$ in conjunction with the enumeration of relatively 
small supercells of various compositions. We validate its screening utility by direct comparison to reported 
experimental $\Delta S_M$ measurements in Mn(Co$_{1-x}$Fe$_x$)Ge and (Mn$_{1-x}$Ni$_x$)CoGe, two systems where 
the magnetocaloric performance depends on $x$ in a highly nonmonotonic manner. In both cases, the method 
successfully predicts the presence and magnitude of enhanced magnetocaloric effects in the solid solutions compared 
to MnCoGe, reproducing the qualitative shape of the $\Delta S_M$ \emph{vs.} $x$ curves and identifying the 
compositions of the largest magnetocaloric effect with errors of $\delta x \approx 0.05$.

\begin{figure}
	\centering
	\includegraphics[width=0.45\textwidth]{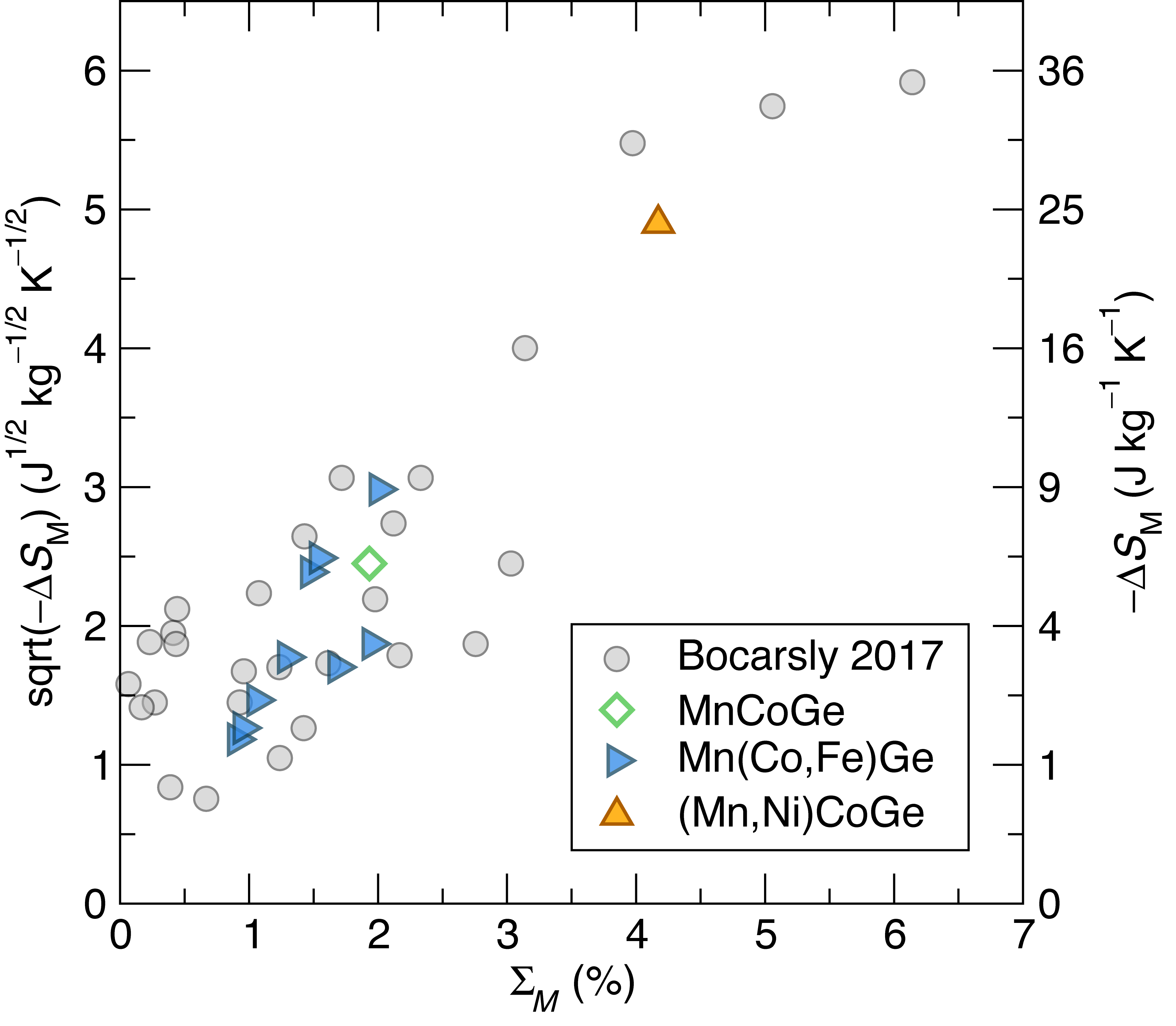}
	\caption{Correspondence between peak values of experimental $-\Delta S_M$ for an applied field of 5\,T and 
	 magnetic deformation $\Sigma_M$. Gray circles show the results from Ref.~\citenum{Bocarsly2017}, which 
	 considered only ordered magnets.
	}  
	\label{fig:summary}
\end{figure}

Fig.~\ref{fig:summary} provides a summary of these results, showing the correspondence between predicted 
$\Sigma_M$ and experimental peak $\Delta S_M$ as compared to previous results applying $\Sigma_M$ to 33 ferromagnets 
without substantial atomic site disorder \cite{Bocarsly2017}. On this plot, all of the calculated $x$ points 
from the Mn(Co$_{1-x}$Fe$_x$)Ge system (Fig.\,\ref{fig:MnCoFeGe_ortho}) are plotted against the $-\Delta S_M$ 
of the nearest composition experimentally reported by Lin \emph{et al.} \cite{Lin2006}. In (Mn$_{1-x}$Ni$_x$)CoGe, 
where the experimental data are more sparse and the composition with maximal $\Sigma_M$ somewhat deviates from the 
reported largest $-\Delta S_M$ \cite{Zhang2015a}, the maximum $\Sigma_M$ (at $x = 0.167$) is plotted against the 
maximum $-\Delta S_M$ (at $x = 0.11$). This plot demonstrates that the magnetocaloric effects of these complex 
disordered systems are being screened with comparable accuracy to prior predictions of ordered magnets.

The success of $\Sigma_M$ in predicting behavior of these complex MnCoGe-based magnetocalorics demonstrates that 
screening disordered magnetocalorics is a promising route towards the discovery of exceptional magnetocaloric 
effects at unstudied compositions. This screening technique requires no information about the system other than 
the known crystal structure of the parent compound (here, MnCoGe), and therefore we believe this approach will be 
quite generally applicable. Due to the many supercells sampled (on the order of 200 for each system), the computational 
cost of obtaining $\Sigma_M$ on solid solutions is much larger than the cost of screening compounds without 
compositional disorder. Nevertheless, it would be feasible to apply this screening method to searches on the order 
of tens, or perhaps hundreds, of systems. For example, this method could be used to exhaustively screen elements 
for promise as dopants for MnCoGe or another magnetocaloric material of interest. 

\section*{Acknowledgments}

This work was supported by the National Science Foundation (NSF) through DMR 1710638.  Use was made of the
computational facilities of the Center for Scientific Computing (CSC) at UC Santa Barbara, supported by
NSF CNS-1725797. The CSC is also supported by the NSF Materials Research Science and Engineering 
Center (DMR 1720256) at UC Santa Barbara. C.A.C.G and J.D.B. have been supported by the National Science 
Foundation Graduate Research Fellowship Program, by Grants DGE-1745303 and DGE-1650114, respectively.


%

\end{document}


\title{Supplemental Material: Computational screening of magnetocaloric alloys}

\author{Christina A. C. Garcia} 
\affiliation{Materials Research Laboratory, University of California, Santa Barbara, California 93106, USA}
\affiliation{John A. Paulson School of Engineering and Applied Sciences, Harvard University, Cambridge, 
Massachusetts 02138, USA}

\author{Joshua D. Bocarsly} 
\affiliation{Materials Research Laboratory, University of California, Santa Barbara, California 93106, USA}
\affiliation{Materials Department, University of California, Santa Barbara, California 93106, USA} 
\email{jdbocarsly@mrl.ucsb.edu}

\author{Ram Seshadri}
\affiliation{Materials Research Laboratory, University of California, Santa Barbara, California 93106, USA}
\affiliation{Materials Department, University of California, Santa Barbara, California 93106, USA} 
\affiliation{Department of Chemistry and Biochemistry, University of California, Santa Barbara, California 93106, USA}

\maketitle

\section{Special Quasi-random Structure calculations}
Special quasirandom structures (SQSs) with different values of $x$ in Mn(Co$_{1-x}$Fe$_x$)Ge (0\,$\leq$\,$x$\,$\leq$\,1) and (Mn$_{1-x}$Ni$_x$)CoGe ($x$\,$\leq$\,0.25) were generated using the \texttt{mcsqs} utility \cite{vandeWalle2013}. An SQS is a single disordered supercell of finite size, constructed such that the local atomic correlation distributions match well those of an infinite randomly distributed alloy. In this case, SQS unit cells containing 48 atoms (four times the primitive orthorhombic TiNiSi-structured cell) were generated to have pair, triplet, and quadruplet correlations of radius $\leq$\,6\,\AA\, that best match the random state. The 48-atom cells allow for compositions with $x$-spacing of 0.0625. 

\begin{figure}[!h]
	\centering
	\includegraphics[width=0.5\columnwidth]{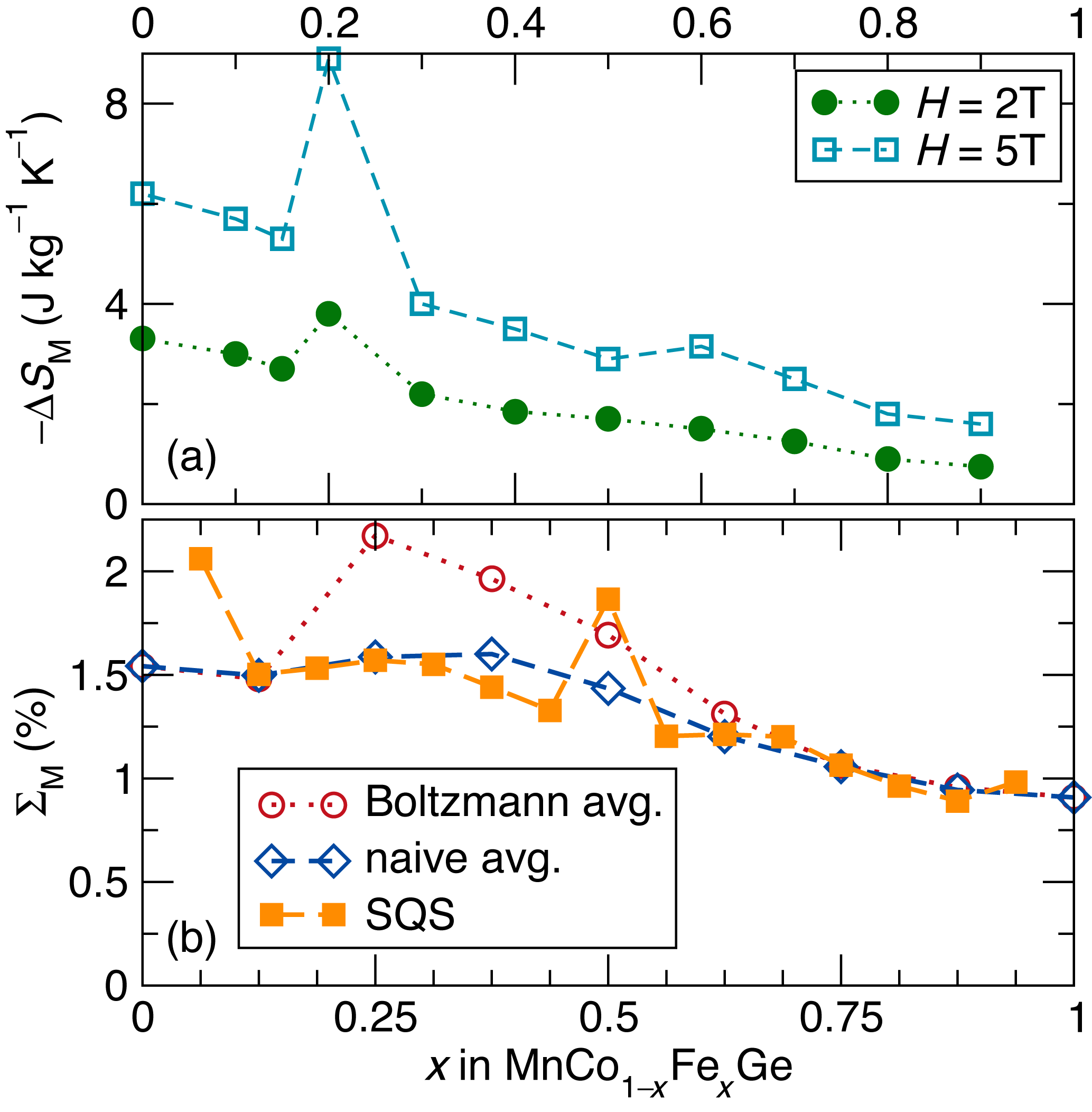}
	\caption{(b) $\Sigma_M$ for Mn(Co$_{1-x}$Fe$_x$)Ge calculated using special quasirandom structures (SQSs) with 48 atoms each is shown compared to $\Sigma_M$ calculating using the supercell approach (main text Fig.~2) with either naive averaging (main text Eqn.~2) or Boltzmann averaging (main text Eqn.~4). (a) shows the experimental peak $\Delta S_M$ for applied fields of 2\,T and 5\,T, as reported by Lin \emph{et al.} \cite{Lin2006}.	The SQS approach gives results similar to the naive average of the supercells, with some deviations.
	}
	\label{fig:MnCoFeGe_sqs}
\end{figure}

As shown in Fig.~\ref{fig:MnCoFeGe_sqs} and Fig.~\ref{fig:MnNiCoFe_sqs}, this approach results in $\Sigma_M$ that closely matches the naive (non-energy-weighted) average of the enumerated supercells, but deviates from the Boltzmann average. This makes sense, as the SQS attempts to reproduce a fully random alloy, with no preference for energetically favorable local configurations. Unfortunately, this means that the SQS does not do a good job of predicting experimental $\Delta S_M$ in this system.

\begin{figure}[!h]
	\centering
	\includegraphics[width=0.5\columnwidth]{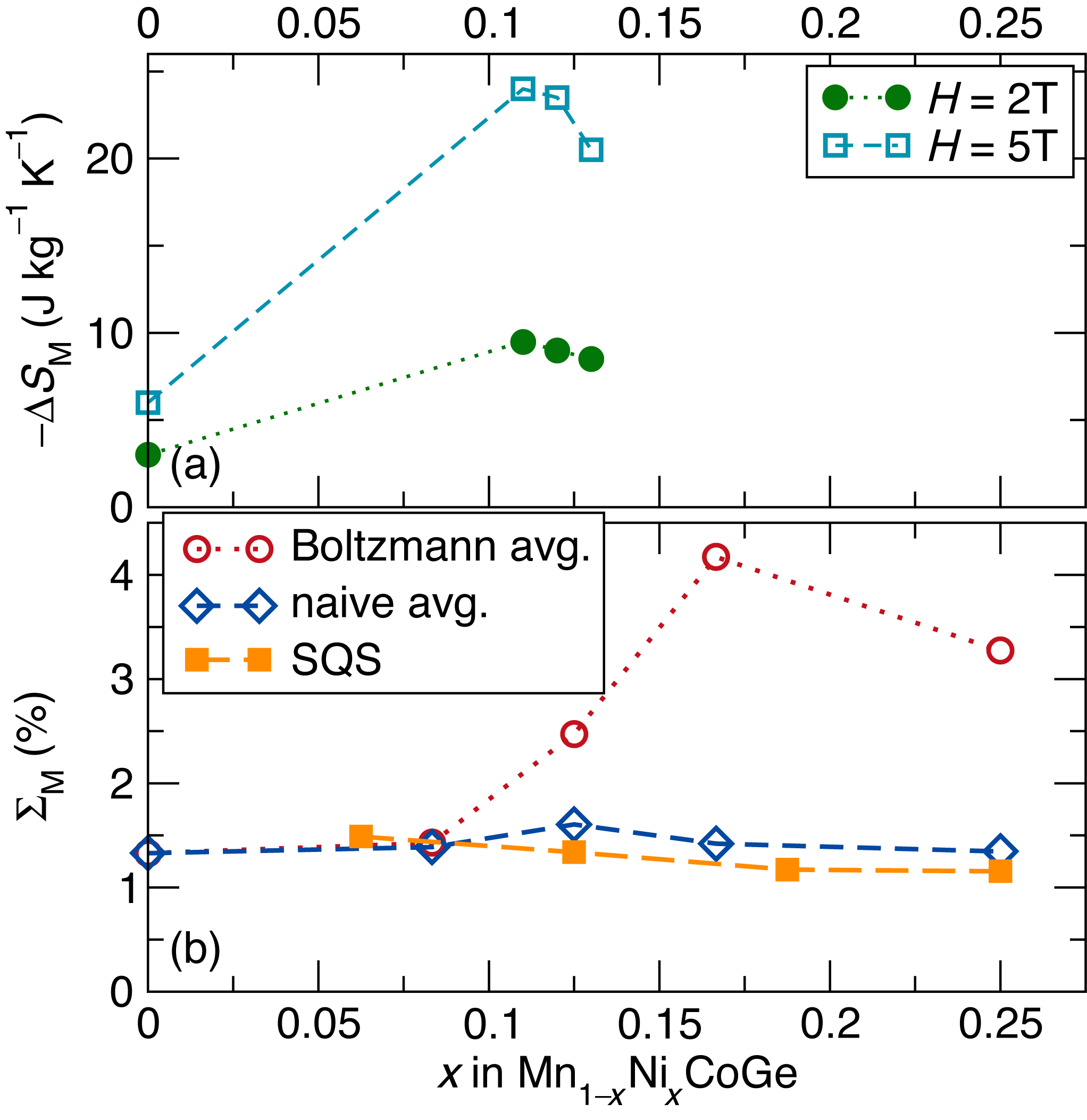}
	\caption{(b) $\Sigma_M$ for (Mn$_{1-x}$Ni$_x$)CoGe ($x \leq 0.25$) calculated using special quasirandom structures (SQSs) with 48 atoms each is shown compared to $\Sigma_M$ calculating using the supercell approach (main text Fig.~2) with either naive averaging (main text Eqn.~2) or Boltzmann averaging (main text Eqn.~4). (a) shows the experimental peak $\Delta S_M$ for applied fields of 2\,T and 5\,T, as reported by Zhang \emph{et al.}\cite{Zhang2015a}. The SQS structure gives results similar to the naive average of the supercells.
	}
	\label{fig:MnNiCoFe_sqs}
\end{figure}

\section{Comparison of $\Sigma_M$ and peak $\Delta S_M$ for M\lowercase{n}(C\lowercase{o}$_{1-x}$F\lowercase{e}$_x$)G\lowercase{e} and (M\lowercase{n}$_{1-x}$N\lowercase{i}$_x$)C\lowercase{o}G\lowercase{e}}

\begin{figure}[!h]
	\centering
	\includegraphics[width=0.75\columnwidth]{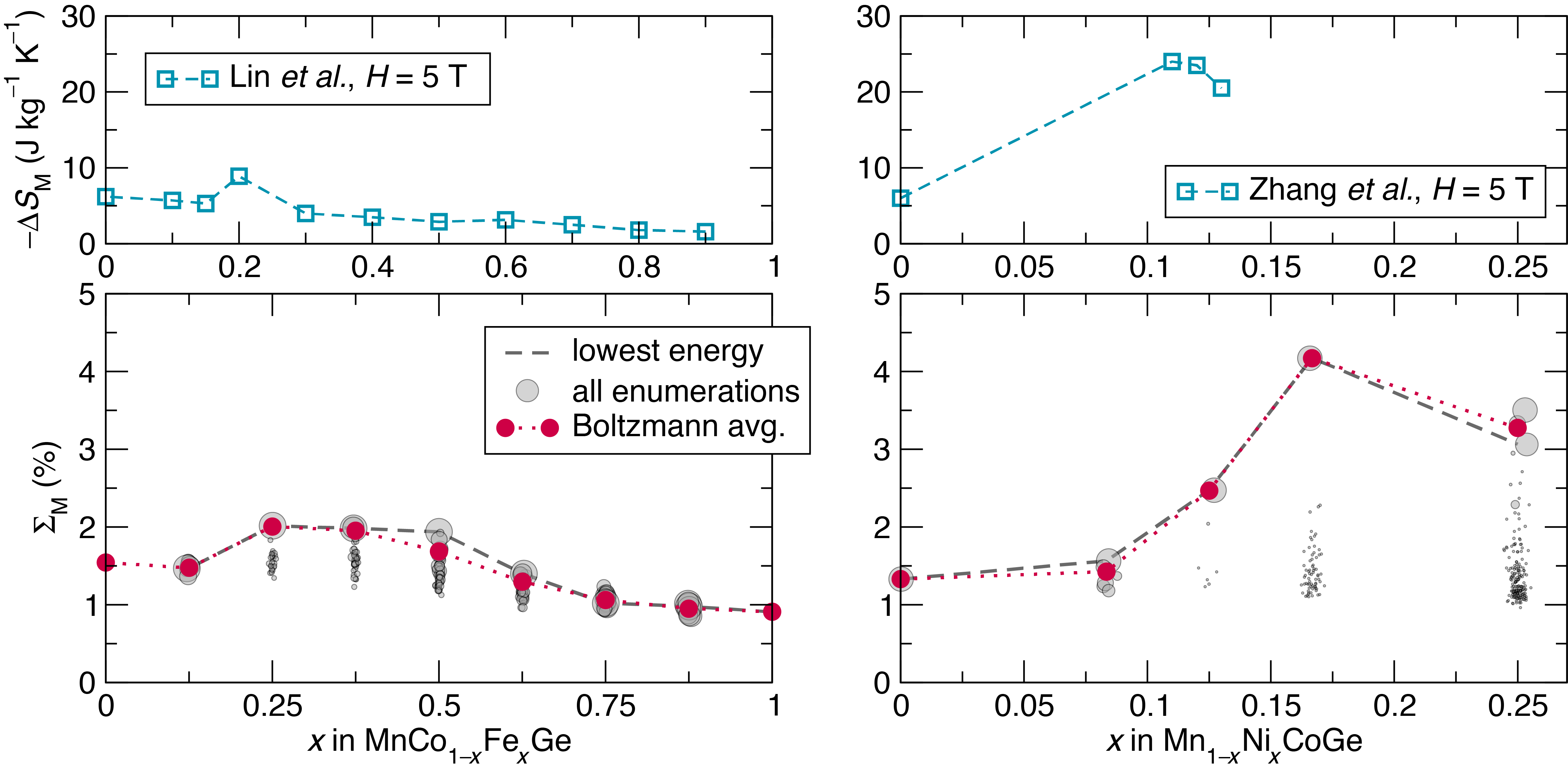}
	\caption{Direct comparison of computed Boltzmann-averaged magnetic deformation ($\Sigma_M$) and reported peak $\Delta S_M$ \cite{Lin2006, Zhang2015a}  for Mn(Co$_{1-x}$Fe$_x$)Ge and (Mn$_{1-x}$Ni$_x$)CoGe ($x \leq 0.25$). This data is reproduced from the main text Fig.~2 and Fig.~3, but presented with the same $y$-axis scaling to show the magnitude differences between the two systems.}
	\label{fig:MnNiCoFe_sqs}
\end{figure}
